\documentclass
[nofootinbib,superscriptaddress,twocolumn,lengthcheck,balancelastpage,oneside,prl]{revtex4}%
\usepackage{amssymb}
\usepackage{amsfonts}
\usepackage{amsmath}
\usepackage{graphicx}
\usepackage{xr}%
\providecommand{\U}[1]{\protect\rule{.1in}{.1in}}

\begin{document}
\title[Interconnected Networks Coupling Threshold]{Exact Coupling Threshold for Structural Transition in Interconnected Networks}
\author{Faryad Darabi Sahneh}
\email{faryad@ksu.edu}
\affiliation{Electrical and Computer Engineering Department, Kansas State University}
\author{Caterina Scoglio}
\affiliation{Electrical and Computer Engineering Department, Kansas State University}
\author{Piet Van Mieghem}
\affiliation{Faculty of Electrical Engineering, Mathematics, and Computer Science, Delft
University of Technology, Delft, The Netherlands}
\keywords{Interconnected Networks, Multiplex, Coupling Threshold, Algebraic Connectivity}
\begin{abstract}
Interconnected networks are mathematical representation of systems where two
or more simple networks are coupled to each other. Depending on the coupling
weight between the two components, the interconnected network can function in
two regimes: one where the two networks are structurally distinguishable, and
one where they are not. The coupling threshold--denoting this structural
transition--is one of the most crucial concepts in interconnected networks.
Yet, current information about the coupling threshold is limited. This letter
presents an analytical expression for the exact value of the coupling
threshold and outlines network interrelation implications.

\end{abstract}
\maketitle

Most natural and human-made networks are not isolated and have external
interactions. Interconnected networks are mathematical representation of
systems where two or more simple networks are coupled to each other. The
importance and challenges of these networks have recently attracted
substantial attention in network science. In particular, researches have
addressed several fundamental problems on dynamical processes over
interconnected networks such as percolation\cite{buldyrev2010N,Hu2011PRE},
epidemic spreading\cite{Saumell2012PRE,HuijuanPRE2013,Sahneh2013ACC}, and
diffusion\cite{Gomez2013PRL}. These networks exhibit properties such as
synchronizability \cite{Aguirre2014PRL}, communicability \cite{Estrada2014PRE}%
, navigability \cite{De2014PNAS}, very different from isolated networks.

Among the most relevant dynamics on networks is the diffusion dynamics.
Hernandez et al. \cite{Hernandez2014PA}\ studied the full spectrum of
interconnected networks where the component networks are identical. Using
perturbation techniques, Gomez et al. \cite{Gomez2013PRL} studied the
diffusion dynamics on interconnected network of two non-identical networks for
weak coupling as well as strong coupling. Significantly, they identified
superdiffusivity, where diffusion in the interconnected network occurs faster
than each network individually. Sole-Ribalta et al. \cite{SoleRibalta2013PRE}
studied the general case, where more than two networks are interconnected with
arbitrary one-to-one correspondence structure. Radicchi and Arenas
\cite{radicchi2013NP} identified a structural transition point depending on
the coupling weight between two networks: the collective interconnected
network can function in two regimes, one where the two networks are
structurally distinguishable and one where they are not. In a similar context,
D'Agostino \cite{Gregorio2014Springer} showed adding intralinks between
networks causes the structural transition from intermode to intramode. For a
class of random network models according to intralayer and interlayer degree
distribution, Radicchi \cite{radicchi2014PRX} showed when correlation between
intralayer and interlayer degrees is below a threshold value, the
interconnected networks become indistinguishable.

\begin{figure}[ptb]%
\centering
\includegraphics[
height=1.4615in,
width=2.5097in
]%
{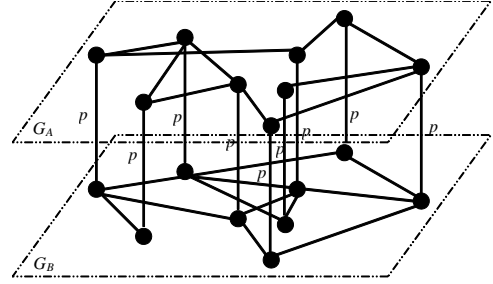}%
\caption{One-to-one interconnection of two networks $G_{A}$ and $G_{B}$, where
the interconnection weight is $p>0$.}%
\label{multiplex.eps}%
\end{figure}

Consider an interconnected network $\boldsymbol{G}$, consisting of two
networks $G_{A}$ and $G_{B}$, each of size $N$, with one-to-one
interconnection with coupling weight $p>0$, as depicted in Figure
\ref{multiplex.eps}. Let matrices $A$ and $B$ represent adjacency matrices of
$G_{A}$ and $G_{B}$, respectively. The overall adjacency matrix and Laplacian
matrix \cite{Piet2011graphspectra} of the interconnected network
$\boldsymbol{G}$ are
\[
\boldsymbol{A}=%
\begin{bmatrix}
A & pI\\
pI & B
\end{bmatrix}
,~\boldsymbol{L}=%
\begin{bmatrix}
L_{A}+pI & -pI\\
-pI & L_{B}+pI
\end{bmatrix}
,
\]
where $L_{A}$ and $L_{B}$ are the Laplacian matrices of $G_{A}$ and $G_{B}$,
respectively, and $I$ is the identity matrix.%

We denote the eigenvalues of the Laplacian matrix $\boldsymbol{L}$\ by
$\,0=\lambda_{1}<\lambda_{2}\leq\cdots\leq\lambda_{2N}$ which satisfy the
following equation%
\begin{equation}%
\begin{bmatrix}
L_{A}+pI & -pI\\
-pI & L_{B}+pI
\end{bmatrix}%
\begin{bmatrix}
V_{A}\\
V_{B}%
\end{bmatrix}
=\lambda%
\begin{bmatrix}
V_{A}\\
V_{B}%
\end{bmatrix}
, \label{Lap_eigs}%
\end{equation}
where $V_{A}$ and $V_{B}$ contain elements of the eigenvector corresponding to
$G_{A}$ and $G_{B}$, respectively, and satisfy the following eigenvector
normalization%
\begin{equation}
V_{A}^{T}V_{A}+V_{B}^{T}V_{B}=2N. \label{V_norm}%
\end{equation}

For the Laplacian matrix $\boldsymbol{L}$, $\lambda_{1}=0$ and the
corresponding eigenvector is $V_{A}=V_{B}=u\triangleq\lbrack1,\ldots,1]^{T}$.
The algebraic connectivity of the interconnected network is the smallest
positive eigenvalue of the Laplacian matrix $\boldsymbol{L}$, which we
represent by $\lambda_{2}(\boldsymbol{L})$. Interestingly, $\lambda=2p$ and
$V_{A}=-V_{B}=u$ is always a solution to the eigenvalue problem
(\ref{Lap_eigs}). Therefore, if $p$ is small enough, the algebraic
connectivity of the interconnected network is $\lambda_{2}=2p$. The
eigenvector corresponding to $\lambda_{2}=2p$, i.e., $V_{A}=-V_{B}=u$,
indicates that networks $G_{A}$ and $G_{B}$ are structurally distinct. By
increasing the coupling weight $p$, this eigenvalue may no longer be the
second smallest one. Recently, Radicchi and Arenas \cite{radicchi2013NP}
argued that there exists a threshold value $p^{\ast}$, so that $\lambda
_{2}=2p$ no longer is the algebraic connectivity for $p>p^{\ast}$. This
transition is an important phenomena as it indicates an abrupt transition in
structure of the interconnected network \cite{radicchi2013NP}: \emph{when
}$p<p^{\ast}$\emph{, the two networks are distinct while for coupling weight
larger than the threshold, the overall interconnected network functions as a
single network}. In other words, for $p>p^{\ast}$, the two networks are not
structurally distinguishable.

Gomez et al. \cite{Gomez2013PRL} showed that the algebraic connectivity of
$\boldsymbol{L}$ is upper-bounded by the half of the algebraic connectivity of
the superpositioned network $G_{s}$ with adjacency matrix $A+B$, i.e.,
$\lambda_{2}(\boldsymbol{L})\leq\frac{1}{2}\lambda_{2}(L_{A}+L_{B})$. This
upper-bound is true for any value of the coupling weight $p$, and becomes
exact as $p\rightarrow\infty$. Using this result, Radicchi and Arenas
\cite{radicchi2013NP} argued that the coupling threshold is upper-bounded by
one fourth of the algebraic connectivity of the super-positioned network,
which is equivalent to%
\begin{equation}
p^{\ast}\leq\frac{1}{2}\lambda_{2}(\frac{L_{A}+L_{B}}{2}). \label{Up_Filippo}%
\end{equation}

Although the coupling threshold $p^{\ast}$ is a critical quantity for
interconnected networks, little is known apart from the upper-bound
(\ref{Up_Filippo}). In this Letter, we \emph{derive the exact value of the
coupling threshold }$p^{\ast}$\emph{ }and present tight bounds that we
interpret physically.

We first need to understand how the eigenvalues of $\boldsymbol{L}$ vary with
$p$. Since the elements of the Laplacian matrix $\boldsymbol{L}$\ are
continuous functions of $p$, so are the eigenvalues of $L$ \cite{zedek1965AMS}%
. This implies that the transition in the algebraic connectivity of the
interconnected network is not a result of any abrupt transitions of the
eigenvalues of $\boldsymbol{L}$, but rather due to crossing of eigenvalues
trajectories as function of $p$. Specifically, the algebraic connectivity
transition occurs precisely at the point where the second and third
eigenvalues of $\boldsymbol{L}$\ coincide. Therefore, \emph{the coupling
threshold }$p^{\ast}$\emph{ is such that }$2p^{\ast}$\emph{ is a repeated
eigenvalue of }$\mathbf{L}$.

Our approach to find the exact value of $p^{\ast}$ is through eigenvalue
sensitivity analysis. The key idea is that while a first-order differentiation
of eigenvalues simply determines eigenvalue/eigenvector sensitivity for
discrete eigenvalues \cite{nelson1976AIAA}, this method cannot uniquely find
the eigen-derivatives for repeated eigenvalues \cite{mills1988AIAA}. Hence, we
study the system of equations for eigenvalue and eigenvector derivatives with
respect to $p$, which we refer to as eigen-derivatives, at $\lambda=2p$, and
look for critical value of $p^{\ast}$ such that a unique solution does not
exist. This Letter includes the main results and procedures of our
mathematical deductions, while further details are available in the
Supplemental Material \cite{Supplement}. Differentiating (\ref{Lap_eigs}) and
(\ref{V_norm}) with respect to $p$ yields the governing equations for the
eigen-derivatives $\frac{dV_{A}}{dp},\frac{dV_{B}}{dp},$ and $\frac{d\lambda
}{dp}$ at $\lambda=2p$%
\begin{equation}%
\begin{bmatrix}
L_{A}-pI & -pI & -u\\
-pI & L_{B}-pI & u\\
-u^{T} & u^{T} & 0
\end{bmatrix}%
\begin{bmatrix}
\frac{dV_{A}}{dp}\\
\frac{dV_{B}}{dp}\\
\frac{d\lambda}{dp}%
\end{bmatrix}
=%
\begin{bmatrix}
-2u\\
2u\\
0
\end{bmatrix}
.\label{Eigen2p}%
\end{equation}
As expected, for $\lambda=2p$ and $V_{A}=-V_{B}=u$, $\frac{dV_{A}}{dp}%
=\frac{dV_{B}}{dp}=0$ and $\frac{d\lambda}{dp}=2$ always satisfy Eq.
(\ref{Eigen2p}). However, the key idea is that when $\lambda=2p$ is a
repetitive eigenvalue, the eigen-derivative equation (\ref{Eigen2p}) does not
have a unique solution. This occurs when the matrix%
\begin{equation}
W\triangleq%
\begin{bmatrix}
L_{A}-pI & -pI & -u\\
-pI & L_{B}-pI & u\\
-u^{T} & u^{T} & 0
\end{bmatrix}
\label{W}%
\end{equation}
is singular. As shown in \cite{Supplement}, $W$ is singular for $p^{\ast
}=\frac{1}{2}\lambda_{i}(Q)$ where the $N\times N$ matrix $Q$ is defined as
$Q\triangleq\bar{L}-\tilde{L}\bar{L}^{\dag}\tilde{L}$, and $\bar{L}$ and
$\tilde{L}$ are%
\begin{equation}
\bar{L}\triangleq\frac{L_{A}+L_{B}}{2},~\tilde{L}\triangleq\frac{L_{A}-L_{B}%
}{2},\label{LtilLbar}%
\end{equation}
and $^{\dag}$ supperscript denotes the Moore--Penrose pseudo-inverse
\cite{Piet2011graphspectra}. Therefore, repeated eigenvalues occur at
$\lambda=2p^{\ast}$ for the values of $p^{\ast}=\frac{1}{2}\lambda_{i}(Q)$,
for $i\in\{1,\cdots,N\}$. This indicates that repeated eigenvalues can occur
for $N$ different values of $p^{\ast}$. For the transition in algebraic
connectivity, the coupling threshold is the smallest positive solution.
Therefore, the exact coupling threshold is%
\begin{equation}
p^{\ast}=\frac{1}{2}\lambda_{2}(Q)\label{p_star}%
\end{equation}

Since term $\tilde{L}\bar{L}^{\dag}\tilde{L}$ in $Q$\ is a positive
semi-definite matrix, $p^{\ast}=\frac{1}{2}\lambda_{2}(Q)=\frac{1}{2}%
\lambda_{2}(\bar{L}-\tilde{L}\bar{L}^{\dag}\tilde{L})\leq\frac{1}{2}%
\lambda_{2}(\bar{L})$ which confirms the upper-bound (\ref{Up_Filippo}) in
\cite{radicchi2013NP}. Interestingly, the exact value not only depends on
$\bar{L}$, half of the Laplacian of the superpositioned network, it also
depends on $\tilde{L}$, which corresponds to the difference between networks
$G_{A}$ and $G_{B}$. After some algebraic manipulations (see \cite{Supplement}%
), $Q$ can be alternatively expressed as%
\begin{align}
Q &  \triangleq\bar{L}-\tilde{L}\bar{L}^{\dag}\tilde{L}\label{Q1}\\
&  =2(L_{A}-\frac{1}{2}L_{A}\bar{L}^{\dag}L_{A})=2(L_{B}-\frac{1}{2}L_{B}%
\bar{L}^{\dag}L_{B})\label{Q2}\\
&  =L_{A}\bar{L}^{\dag}L_{B}=L_{B}\bar{L}^{\dag}L_{A}\label{Q3}%
\end{align}
Furthermore, according to (\ref{p_star}) and (\ref{Q3}), the coupling
threshold $p^{\ast}$ can be alternatively obtained as%
\begin{equation}
p^{\ast}=\frac{1}{\rho(L_{A}^{\dag}+L_{B}^{\dag})},\label{pstar_ind}%
\end{equation}
where $\rho(\bullet)\triangleq\lambda_{N}(\bullet)$ denotes spectral radius
(see \cite{Supplement}). Finally, expressions (\ref{Q1}), (\ref{Q2}), and
(\ref{Q3}) for $Q$ provide upper-bound and lower bound for the coupling
threshold $p^{\ast}=\frac{1}{2}\lambda_{2}(Q)$ in terms of the spectral radius
of each isolated network $G_{A}$ and $G_{B}$, as well as the super-positioned
network $G_{s}$ as
\begin{align}
p^{\ast} &  \geq\frac{1}{\lambda_{2}^{-1}(L_{A})+\lambda_{2}^{-1}(L_{B}%
)},\label{Low1}\\
p^{\ast} &  \leq\min\{\lambda_{2}(L_{A}),\lambda_{2}(L_{B}),\frac{1}{2}%
\lambda_{2}(\bar{L})\}.\label{Up1}%
\end{align}

The lower-bound (\ref{Low1}) has a very elegant expression, as it is half of
the \emph{harmonic mean} of $\lambda_{2}(L_{A})$ and $\lambda_{2}(L_{B})$. The
upper-bounds (\ref{Up1}) not only includes the upper-bound $\frac{1}{2}%
\lambda_{2}(\bar{L})$, reported in \cite{radicchi2013NP}, but also it
indicates a fundamental property of interconnected networks: the coupling
threshold $p^{\ast}$ is upper-bounded by the algebraic connectivity of the
least-connected network. Furthermore, if the algebraic connectivity of one
network is at least three times smaller than that of the other network, i.e.,
$\lambda_{2}(L_{A})<\frac{1}{3}\lambda_{2}(L_{B})$ without loss of generality,
then the algebraic connectivity of the least-connected network, here $G_{A}$,
mainly determines the coupling threshold, and the super-positioned network
does not play a major role. Indeed, if $K\triangleq\lambda_{2}(L_{B}%
)/\lambda_{2}(L_{A})>3$, then%
\begin{equation}
\frac{K}{1+K}\lambda_{2}(L_{A})<p^{\ast}\leq\lambda_{2}(L_{A}%
).\label{lam2sandwich}%
\end{equation}

While the upper-bounds and lower-bound (\ref{Low1}) and (\ref{Up1}) are
simple, they do not include much information regarding interrelations of
network components. We can find bounds that explicitly depend on the networks
interrelations. We can use formula (\ref{pstar_ind}) to build an upper-bound
$p^{\ast}\leq\frac{1}{\hat{\rho}_{n_{A},n_{B}}}$\ using the eigenvectors
corresponding to the $n_{A}$ smallest positive eigenvalue of $L_{A}$ and the
$n_{B}$ smallest positive eigenvalue of $L_{B}$ \cite{Supplement}, where
$\hat{\rho}_{n_{A},n_{B}}$ is the spectral radius of an $(n_{A}+n_{B}%
)-$by$-(n_{A}+n_{B})$ matrix, i.e.,\vskip-\parskip{\footnotesize
\begin{equation}
\hat{\rho}_{n_{A},n_{B}}=\rho\left(  u_{(n_{A}+n_{B})}%
\begin{bmatrix}
\boldsymbol{\lambda}^{-1}(L_{A})\\
\boldsymbol{\lambda}^{-1}(L_{B})
\end{bmatrix}
^{T}\circ%
\begin{bmatrix}
I_{n_{A}} & \boldsymbol{v}_{A}^{T}\boldsymbol{v}_{B}\\
\boldsymbol{v}_{B}^{T}\boldsymbol{v}_{A} & I_{n_{B}}%
\end{bmatrix}
\right)  , \label{RhonAnB}%
\end{equation}
} where, $\circ$ denotes the Hadamard (entry-wise) product,
$\boldsymbol{\lambda}^{-1}(L_{A})\triangleq\lbrack\lambda_{2}^{-1}%
(L_{A}),\cdots,\lambda_{n_{A}+1}^{-1}(L_{A})]^{T}$, $\boldsymbol{v}_{A}%
=[v_{2}(L_{A}),\cdots,v_{n_{A}+1}(L_{A})]\in%
\mathbb{R}
^{N\times n_{A}}$, and $\boldsymbol{\lambda}^{-1}(L_{B})$ and $\boldsymbol{v}%
_{B}$ are defined similarly. The interesting aspect of this upper-bound is
that it not only depends on the smallest positive eigenvalues of $L_{A}$ and
$L_{B}$, it also depends on the inner-product of their corresponding
eigenvectors, thus explicitly incorporating networks interrelation. By
computing a few eigenvectors of $L_{A}$ and $L_{B}$, this upper-bound gives
very good estimates, with increasing precision as the number of eigenvectors
$n_{A}$ and $n_{B}$ increases.

\begin{figure}[ptb]%
\centering
\includegraphics[
height=2.3609in,
width=3.0113in
]%
{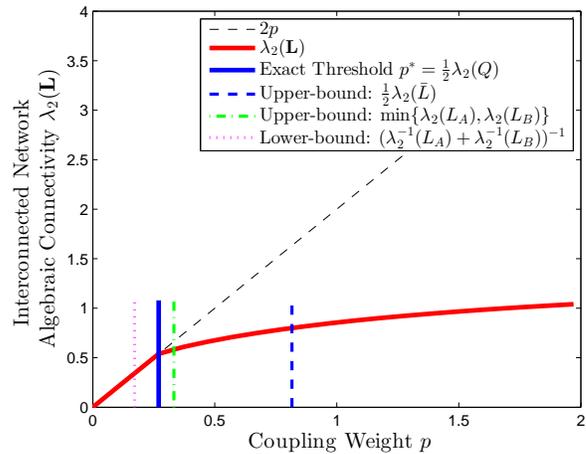}%
\caption{Algebraic connectivity $\lambda_{2}(\boldsymbol{L})$ of two coupled
networks as a function of the coupling weight $p$. For $p<p^{\ast}=0.27\,,$
algebraic connectivity is $\lambda_{2}(\boldsymbol{L})=2p$. For $p>p^{\ast}$,
eigenvalue $\lambda=2p$ is no longer the algebraic connectivity of the
interconnected network; thus, denoting a structural transition at $p=p^{\ast}%
$. }%
\label{couplingthreshold.eps}%
\end{figure}

In the following, we perform several numerical simulations to investigate our
analytical results. First, we generate an interconnected network with
$N=1000$, where graph $G_{A}$ is a scale-free network with exponent $\gamma
=3$, and $G_{B}$ is a random geometric network with threshold distance
$r_{c}=\sqrt{\frac{5\log N}{\pi N}}$. For these networks, $\lambda_{2}%
(L_{A})=0.355$, and $\lambda_{2}(L_{B})=0.332$. Figure
\ref{couplingthreshold.eps} shows the algebraic connectivity $\lambda_{2}(L)$
of the interconnected network as a function of the coupling weight $p$, and
illustrates that formula (\ref{p_star}) predicts the coupling threshold
exactly. Furthermore, this simulation supports the analytical results for
bounds in (\ref{Up1}) and (\ref{Low1}). In order to highlight different
aspects of topological properties of interconnected networks, we design two
numerical experiments: one for a set of interconnected networks
$\boldsymbol{G}$ with identical superpositioned network $G_{s}$, and one for a
set of interconnected networks $\boldsymbol{G}$ with isomorphic network
components $G_{A}$ and $G_{B}$.%

\begin{figure}[ptb]%
\centering
\includegraphics[
height=2.2494in,
width=3.0113in
]%
{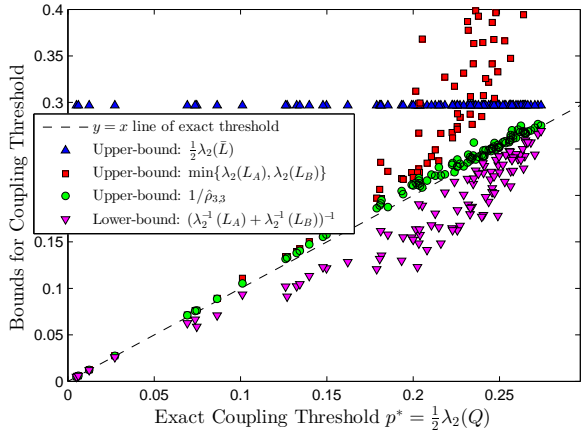}%
\caption{Bounds for the coupling threshold versus the exact values for a set
of interconnected networks with identical averaged network. Specifically,
weighted networks $G_{A}$ and $G_{B}$ are randomly generated such that $A+B$
is the adjacency matrix of the weighted Karate Club network. For each
generated network, we compute different bounds for the coupling threshold and
compare them with the exact value. The closer to the black dashed line, the
more accurate the bounds.}%
\label{karateclub_identicalaverage.eps}%
\end{figure}

For the first set of interconnected networks with identical superpositioned
network, we generate a set of interconnected networks from the Karate Club
network according to the following rule: $a_{ij}=a_{ji}=p_{ij}w_{ij}$ and
$b_{ij}=b_{ji}=(1-p_{ij})w_{ij}$ for $j<i$, where $w_{ij}$'s are the elements
of the weighted Karate Club adjacency matrix and $p_{ij}$'s are i.i.d.
uniformly distributed on $\left[  0,1\right]  $. In this way, the
super-positioned network will always be the same for any realization of this
interconnected network generation. Therefore, differences in the outputs do
not depend on the superpositioned network. Figure
\ref{karateclub_identicalaverage.eps} shows different bounds for the coupling
threshold versus the exact values. The upper-bound $\frac{1}{2}\lambda
_{2}(\bar{L})$ is the same even though the exact threshold $p^{\ast}$ has a
broad distribution. When $p^{\ast}$ is small, the upper-bound $\min
\{\lambda_{2}(A)$,$\lambda_{2}(B)\}$ is accurate, i.e. close to, but above the
$y=x$ line (black dashed line). This region represents interdependent networks
where one network component is loosely connected and possesses a relatively
small algebraic connectivity. As discussed in (\ref{lam2sandwich}), in these
cases the value of the coupling threshold is mainly determined by the
algebraic connectivity of the least connected network, which explains why
$\min\{\lambda_{2}(A)$,$\lambda_{2}(B)\}$ leads to accurate predictions.%

\begin{figure}[ptb]%
\centering
\includegraphics[
height=2.2909in,
width=3.0113in
]%
{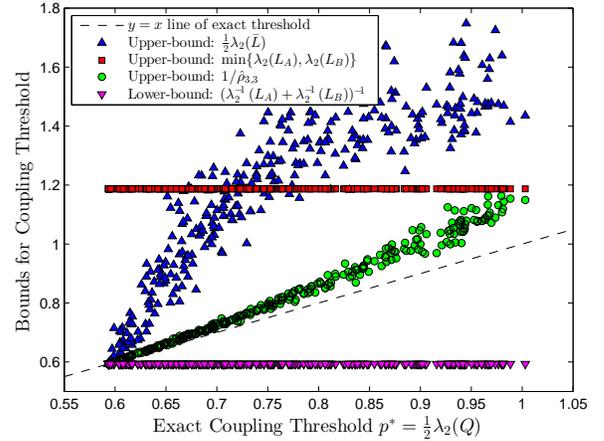}%
\caption{Bounds for the coupling threshold versus the exact values for a set
of interconnected networks where network $G_{A}$ and $G_{B}$ are isomorphic
and thus have identical graph properties. For each generated network, we
compute different bounds for the coupling threshold and compare them with the
exact value. The closer to the black dashed line, the more accurate the
bounds.}%
\label{karateclub_identicallayers.eps}%
\end{figure}

For the second set of interconnected networks with isomorphic network
components $G_{A}$ and $G_{B}$, we generate another set of interconnected
networks for which we use the adjacency matrix of the Karate club network as
$A$, and then pick the adjacency matrix of $G_{B}$ as $B=P^{-1}AP$, where $P$
is a randomly chosen permutation matrix. In this way, $G_{B}$ is basically the
Karate Club network, however, with different node labels. Therefore, $G_{A}$
and $G_{B}$ are isomorphic and have identical graph properties. Therefore,
different outputs are purely due to the interrelation between $G_{A}$ and
$G_{B}$. For each generation of such interconnected network, Figure
\ref{karateclub_identicallayers.eps} shows several bounds for the coupling
threshold plotted versus the exact value. Note that the upper-bound
$\min\{\lambda_{2}(A)$,$\lambda_{2}(B)\}$ and lower-bound $(\lambda_{2}%
^{-1}(A)+\lambda_{2}^{-1}(B))^{-1}$ are always constant, as these values only
depend on the graph properties of $G_{A}$ and $G_{B}$, which are kept
identical. There is a significant negative correlation between the coupling
threshold and Fiedler vectors of $G_{A}$ and $G_{B}$ (i.e., $|v_{2}^{T}%
(L_{A})v_{2}(L_{B})|$). The coupling threshold is maximal when the two
networks are uncorrelated (i.e., $|v_{2}^{T}(L_{A})v_{2}(L_{B})|\rightarrow0$)
and decreases as the two networks become more correlated ($|v_{2}^{T}%
(L_{A})v_{2}(L_{B})|\rightarrow1$). We remark that here the correlation
between $G_{A}$ and $G_{B}$ is measured in terms of their Fiedler vectors, and
that other correlation metrics--such as degree correlation-- do not
necessarily yield similar results. See, \cite{Supplement} for further
information.%

In conclusion, this Letter computes exactly the critical value $p^{\ast}$ for
the coupling weight in an interconnected network $\boldsymbol{G}$, for which
only a few bounds were known so far. The exact expression of the coupling
threshold $p^{\ast}$\ not only depends on individual network components
$G_{A}$ and $G_{B}$ or the superpositioned network $G_{s}$, but also depends
on the interrelation of $G_{A}$ and $G_{B}$. Yet, it is possible to detect
upper and lower bounds for the coupling threshold $p^{\ast}$ only in terms
graph properties of $G_{A}$, $G_{B}$, and $G_{s}$. These types of bounds are
important, even though they lack a description of the interconnection relation
between $G_{A}$ and $G_{B}$. The exact expression for $p^{\ast}$ directly led
to new upper and lower bounds only in terms of graph properties of $G_{A}$,
$G_{B}$, and $G_{s}$. Furthermore, we developed the upper-bound (\ref{RhonAnB}%
) with tunable accuracy, which explicitly depends on the network
interrelation. Through analytic arguments and a specific design of numerical
experiments, we showed that the superpositioned network $G_{s}$ is physically
irrelevant for the identification of the coupling threshold when one of the
network components is considerably less connected, or when the network
components $G_{A}$ and $G_{B}$ are uncorrelated according to their Fiedler
eigenvectors inner product, i.e., when $|v_{2}^{T}(L_{A})v_{2}(L_{B})|$ is
small. Even though the analysis has been performed for coupling of two
networks, we expect the methodology to be generalizable to multiple coupled
networks, as $p^{\ast}$ is the critical value for the coupling weight $p$ for
which the eigen-derivative equations do not have unique solutions. Hence, this
Letter sheds new light on the true nature of structural transitions in
interconnected networks, outlining the importance of topological
interrelations in such networks.

\textbf{Acknowledgement.} We would like to thank Filippo Radicchi and Alex
Arenas for their helpful suggestions to improve this manuscript. This work has
been supported by the National Science Foundation Award CIF-1423411. Any
findings, recommendations, and opinions in this work are those of the authors
and do not necessarily reflect the views of the National Science Foundation.

\bibliographystyle{apsrev}
\bibliography{MultiplexTrans}

\end{document}